\documentclass[aps,twocolumn,prl,preprintnumbers,amsmath,amssymb]{revtex4}
\usepackage{graphicx, bm}
\usepackage{dcolumn}
\usepackage{float}
\usepackage{latexsym}
\usepackage{amsmath}
\usepackage{graphics}
\usepackage{amssymb}
\usepackage{layout}
\usepackage{verbatim}
\usepackage{amsfonts,epsfig}
\usepackage{color}
\usepackage{setspace}
\newcommand{\beqnar}{\begin{eqnarray}}
\newcommand{\eeqnar}{\end{eqnarray}}

\newcommand{\bq}{{\bf q }}
\newcommand{\br}{{\bf r }}

\newcommand{\beq}{\begin{equation}}
\newcommand{\eeq}{\end{equation}}
\newcommand{\nav}    {\langle n\rangle}
\newcommand{\nrms}   {n_{\rm rms}}

\begin{document}
\title{Theory of 2D transport in graphene for correlated disorder}
\author{Qiuzi Li$^1$, E. H.\ Hwang$^1$, E. Rossi$^2$, and S. Das Sarma$^1$}
\affiliation{$^1$Condensed Matter Theory Center, Department of Physics, University of Maryland, College Park, Maryland 20742\\
             $^2$Department of Physics, College of William and Mary, Williamsburg, VA 23187, USA}
\date{\today}
\begin{abstract}
We theoretically revisit graphene transport properties as a function
of carrier density, taking into
account possible correlations in the spatial distribution of the
Coulomb impurity disorder in the environment. We find that the charged
impurity correlations give rise to a
density dependent graphene conductivity, which agrees well
qualitatively with the existing experimental data.
We also find, quite
unexpectedly, that the conductivity could increase with increasing
impurity density if there is sufficient inter-impurity correlation
present in the system.  In
particular, the linearity (sublinearity) of graphene conductivity at
lower (higher) gate voltage is naturally explained as arising solely
from impurity correlation effects in the Coulomb disorder.
\end{abstract}


\maketitle

One of the most studied properties of graphene is its electrical
conductivity as a function of the applied gate voltage
which translates directly
into the carrier density $(n)$ dependent conductivity $\sigma(n)$
\cite{dassarma2010}.
The functional dependence of $\sigma(n)$  at low temperatures
contains information \cite{dassarma2010} about the nature of disorder
in the graphene environment giving rise to the dominant resistive
carrier scattering mechanism. Although there is a well-accepted theory
\cite{dassarma2010} for graphene transport involving an interplay
between long-range charged impurity and short-range disorder
scattering, the theory is not universally accepted and cannot explain
all experimental observations, indicating the possibility of important
missing ingredients \cite{katsnelson2008}.

In this work, we provide a qualitatively new theory for the
$\sigma(n)$ properties of graphene and introduce a new physical
explanation for the experimental observations, i.e., we explain why
$\sigma(n)\sim n$ for `small' or `intermediate' $n$ and
$\sigma(n)\sim$ constant for `large' $n$, with a smooth nonlinear
crossover between the two asymptotic behaviors.
We also provide theoretical results
for $\sigma_{min}$, the graphene minimum conductivity at the Dirac point, using our new theory.
We concentrate on the nature of the underlying
static disorder limiting graphene transport in currently available
samples where phonon scattering effects are relatively weak (compared
with disorder scattering) even at room temperatures
\cite{EfetovKim_PRL10}. The quantitative
weakness of the electron-phonon interaction in graphene gives
particular impetus to a thorough understanding of the disorder
mechanisms limiting graphene conductivity since this may enable
substantial enhancement of room temperature graphene-based device
speed for technological applications as disorder remains the primary
resistive mechanism limiting graphene transport even at room
temperatures.
Therefore, a complete understanding of
the disorder mechanisms controlling $\sigma(n)$ in graphene at $T=0$
is of utmost importance both from fundamental and technological
prospectives.

The most important
features of the experimentally observed $\sigma(n)$ \cite{Novoselov,TanDas_PRL07,ChenJ_NPH_2007,BolotinYacoby,ZhuExp_PRB09} in
graphene are: (1) a nonuniversal sample-dependent minimum conductivity
$\sigma(n\approx0)\equiv\sigma_{min}$ at the charge neutrality point
(CNP) where the average carrier density vanishes;
(2) a linearly increasing,
$\sigma(n)\propto n$ , conductivity with increasing carrier density on
both sides of the CNP upto some sample dependent characteristic
carrier density; (3) a sublinear $\sigma(n)$ for high carrier density,
making it appear that the very high density $\sigma(n)$ may be
saturating.

A successful model
\cite{dassarma2010,Adam, Rossi,HwangAdamDas_PRL07,AndoMac}
for diffusive graphene carrier transport incorporates two
distinct scattering mechanisms with individual resistivity $\rho_c$
and $\rho_s$, arising respectively from the long-range Coulomb
disorder due to random background charged impurities and static
zero-range (often called ``short-range") disorder. The net graphene
conductivity is then given by $\sigma\equiv\rho^{-1}=(\rho_c +
\rho_s)^{-1}$. It is easy to show that
\cite{dassarma2010,Adam,Rossi,HwangAdamDas_PRL07,AndoMac}
$\rho_c\sim1/n$ and $\rho_s\sim$ constant in graphene, leading to
$\sigma(n)$ going as
$\sigma(n) = {n}/({A + C n})$
where the constants $A$ and $C$ are known \cite{dassarma2010} as
functions of disorder parameters; $A$, arising from Coulomb disorder,
depends on the impurity density ($n_i$)
(and also on their locations in space)
and the background dielectric constant ($\kappa$)
whereas the constant $C$, arising from the short-range disorder
\cite{dassarma2010,HwangAdamDas_PRL07}, depends on the strength of the
white-noise disorder characterizing the zero-range scattering.
The relation $\sigma(n) = {n}/({A + C n})$
explains the observed $\sigma(n)$ behavior
of graphene for $n\neq 0$ since $\sigma(n\ll A/C)\sim n$, and
$\sigma(n\gg A/C)\sim1/C$ with $\sigma(n)$ showing sublinear
$(C+A/n)^{-1}$ behavior for $n\sim A/C$.

The above-discussed scenario for disorder-limited graphene
conductivity, with both long-range and short-range disorder playing
important qualitative roles at intermediate $(n_i\lesssim n\leqslant
A/C)$ and high $(n>A/C)$ carrier densities respectively, has been
experimentally verified by several groups
\cite{TanDas_PRL07,ChenJ_NPH_2007,BolotinYacoby,ZhuExp_PRB09}.
%
%
There is, however, one {\it serious issue } with this reasonable scenario:
%
although the physical mechanism underlying the long-range
disorder scattering is experimentally established \cite{dassarma2010,TanDas_PRL07,ChenJ_NPH_2007} to be the presence of
unintentional charged impurity centers in the graphene environment,
the physical origin of the short-range disorder scattering is unclear
and experimentally obscure. Point defects (e.g. vacancies) are rare
in graphene producing negligible short-range disorder. There have also been
occasional puzzling conductivity measurements
[e.g., Ref.~\onlinecite{GeimIce_PRL09}] reported in the literature which
do not appear to be easily explicable using the standard model of
independent dual scattering by long- and short-range disorder playing
equivalent roles.

In this Letter we propose an alternative physical model for
understanding disorder-limited $\sigma(n)$ behavior in graphene. The
model is simpler (and therefore, more appealing) than the standard
model of independent dual disorder mechanisms because it requires only
the long-range Coulomb disorder associated with the background charged
impurities eliminating completely the {\it ad hoc} short-range
disorder necessary  for explaining the
high-density nonlinearity in $\sigma(n)$. Our model, therefore,
eliminates the undesirable feature of the standard model, namely, no
adjustable short-range scattering term with unknown physical origin
needs to be arbitrarily added to the problem in order to explain the
observed high-density sublinear $\sigma(n)$.

The key to our model is the inclusion of some {\it spatial
  correlations} in the distribution of the charged impurity locations
in the system, i.e., the charged impurities are no longer considered
to be completely random spatially. Some impurity correlations are
perfectly reasonable to assume since much of the fabrication and
processing of graphene is done at room temperature (and in fact, often
thermal and/or current annealing is used in sample preparation), which
is expected to lead to actual diffusion of the impurities producing an
annealed, at least partially, correlated impurity configuration rather
than a quenched uncorrelated random one.
We show that the single assumption of impurity
correlations, defined through a correlation length scale parameter
$r_0$, is sufficient to explain the qualitative features of the
experimental $\sigma(n)$ behavior using only disorder scattering  by
background charged impurities.

\begin{figure}
\includegraphics[width=0.99\columnwidth]{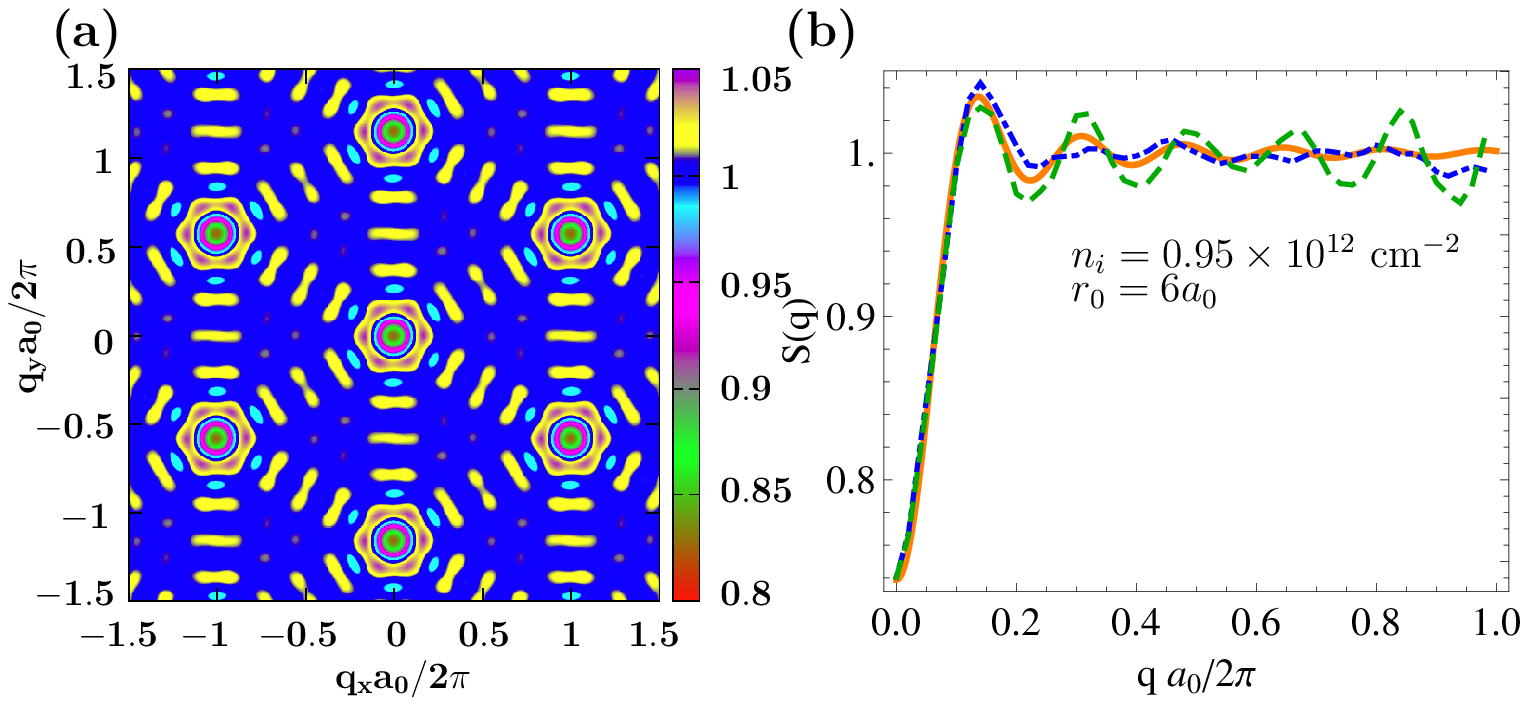}
\caption{
 (a) Density plot of structure factor $S(\bq)$ obtained from Monte
  Carlo simulations
  for $n_i = 0.95\times10^{12}$~cm$^{-2}$, $a_0=4.92\;\AA$ and $r_0=5a_0$.
  (b) Structure factor $S(\bq)$ using Eq.~(\ref{eq:strufac2}) (solid line)
and Monte Carlo simulations.
  Dot-dashed and dashed lines show the Monte Carlo results
  for two different directions of $\bq$ from
  $x$-axis, $\theta = 0$ and
  $\theta=30^{\circ}$, respectively.
}
\label{fig:1}
\end{figure}

To calculate the impurity correlations we use Monte Carlo simulations
carried out on a $200\times200$ triangular lattice with $10^6$ averaging runs,
periodic boundary conditions, and a lattice constant
$a_0=4.92${\AA} which is two times the graphene lattice constant since
the most closely packed phase of impurity atoms (e.g. K as in
Ref. \onlinecite{ChenJ_NPH_2007}) on graphene is likely to be an
$m\times m$ phase with $m=2$ for K \cite{Caragiu_JPCM05}.
%
Correlations are
automatically introduced by virtue of the random positioning of the
impurities at lattice sites with the correlation length $r_0<r_i=(\pi n_i)^{-1/2}$.
Our correlation
model is physically motivated with the reasonable underlying
assumption that two impurities cannot be arbitrarily close to each
other (as they can be in the unphysical continuum random impurity
model, where $r_0=0$), and there must be a minimum separation between
them.
A reasonable continuum approximation to this discrete lattice model
is given by the following pair distribution function $g({\bf r})$
(${\bf r}$ is a 2D vector in the graphene plane)
\begin{align}
g(\br) =\begin{cases}
0 & |\br| \leq r_0\\
1 & |\br| > r_0 \end{cases}.
\label{eq:g}\end{align}
for the impurity density distribution.
Even though Eq.~(\ref{eq:g}) is only an approximation
the basic idea of a length scale $r_0$ defining the
spatial impurity correlations is physically sound (with $r_0=0$ for
the purely random case).
Impurity correlation
effects enter the transport theory through the structure factor $S(\bq)$,
given by
$S(\bq)  =1+n_{i}\int d^{2}r e^{i\mathbf{q\cdot
    r}}[g(\mathbf{r})-1]$.
For uncorrelated random impurity scattering, as in the standard theory,
$g(\br)=1$ always, and $S(\bq)\equiv 1$.
With Eq.~(\ref{eq:g}), we have
\begin{equation}
S(q) =1-2\pi n_{i}\frac{r_{0}}{q}J_{1}(qr_{0})
\label{eq:strufac2}
\end{equation}
where $J_1(x)$ is the Bessel function of the first kind.
Fig.~\ref{fig:1} (a) shows the
structure factor $S(\bq)$ obtained from the
Monte Carlo simulations.
Fig. \ref{fig:1}(b) shows
$S(\bq)$ for both the random Monte Carlo realistic numerical model and
the simple continuum analytic approximation [Eq.~(\ref{eq:strufac2})]. It is obvious that the
analytic approximation captures well the essential
features of the full numerical Monte Carlo simulation.

The graphene carrier
conductivity due to scattering by screened Coulomb disorder can now be
calculated taking into account the impurity correlations, leading to
$ \sigma=({e^2}/{h})({gE_F\tau})/({2\hbar})$,
%
where $E_F$ is the Fermi
energy, $g=4$ is the total degeneracy of
graphene, and the transport relaxation time
$\tau$ is given by, \cite{HwangDas_PRB07}
\begin{equation}
\dfrac{\hbar}{\tau}=\Bigg(\dfrac{\pi n_{i}\text{\ensuremath{\hbar}}
v_{F}}{4k_{F}}\Bigg)r_{s}^{2}\int
\frac{d\theta \left(1-\cos^{2}\theta\right)}
{\left(\sin\frac{\theta}{2}+2r_{s}\right)^2}
S(2k_{F}\sin\frac{\theta}{2}),
\label{eq:relatime}
\end{equation}
where $v_F$ is graphene Fermi velocity,
$k_F$ the Fermi
wavevector ($k_F=E_F/(\hbar v_F)$), and $r_s$ the graphene fine structure
constant ($r_s=e^2/(\hbar v_F\kappa)$).
%
For uncorrelated random impurity scattering,
$r_0 = 0$, $g(\br)=1$, $S(\bq)\equiv 1$, we recover
%
the standard formula for Boltzmann conductivity by screened random charged
impurity centers \cite{HwangAdamDas_PRL07,AndoMac}.
%
%
\begin{figure}
\includegraphics[width=0.99\columnwidth]{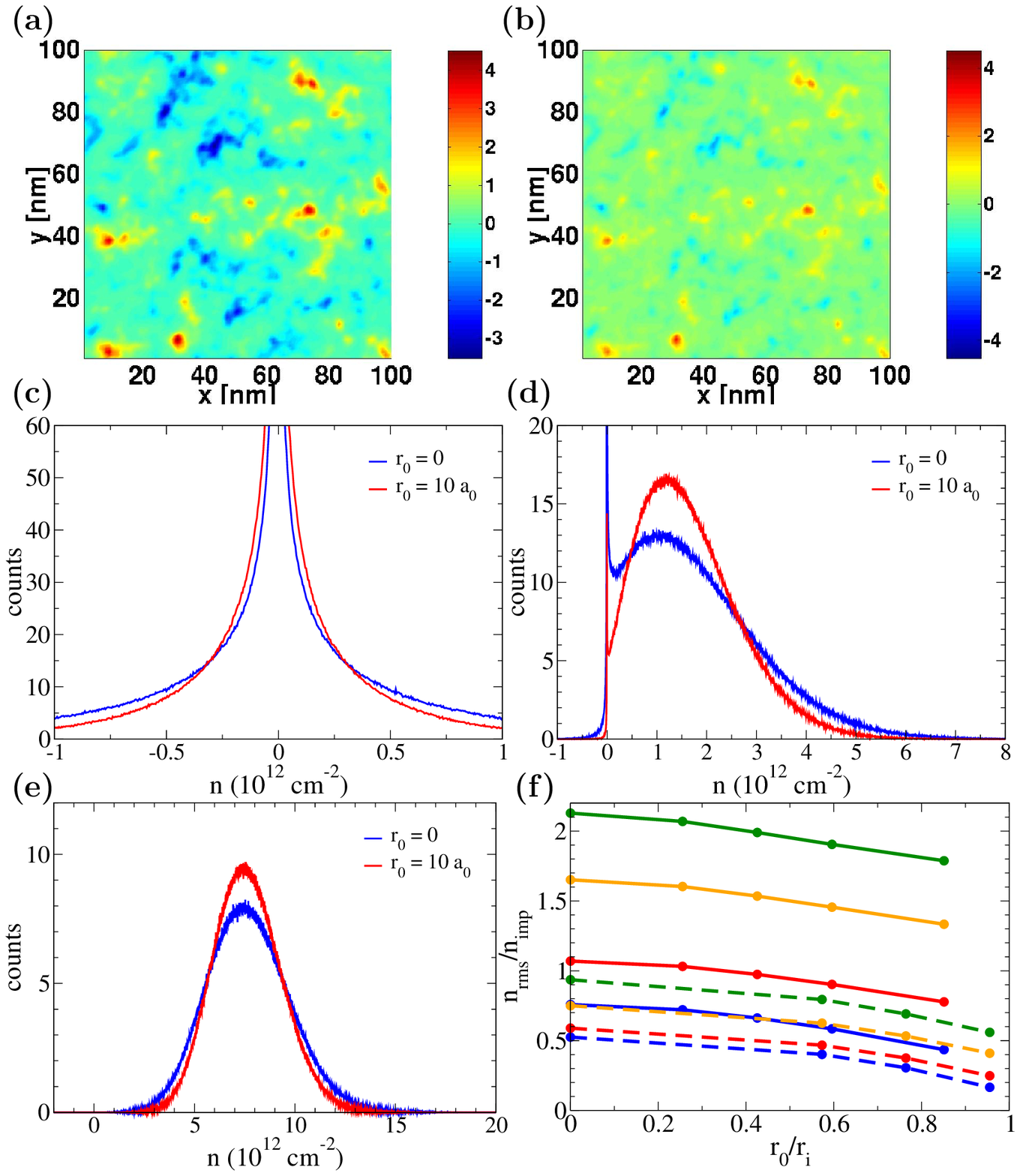}
\caption{(color online)
  The carrier density for a single disorder
  realization obtained from the TFD theory (a)
  for the uncorrelated case and (b) $r_0=10\;a_0$ with  $n_i
  =0.95\times10^{12}$~cm$^{-2}$.
  Carrier probability distribution function $P(n)$
are  shown in (c), (d), (e) for
  $\nav=0$, 1.78, $7.7\times 10^{12}$~cm$^{-2}$, respectively.
  In (f) the ratio $n_{\rm rms}/n_i$ is shown as a function of $r_0/r_i$
  for $n_i = 0.95\times10^{12}$ cm$^{-2}$, solid lines, and
  $n_i = 4.8\times10^{12}$ cm$^{-2}$, dashed lines.
We use $\nav=7.7$, 3.14, 0.94, $0\times 10^{12}$~cm$^{-2}$
  for the solid lines (from top to bottom) and
  $\nav=8.34$, 4.10, 1.7, $0\times 10^{12}$~cm$^{-2}$
  for the dashed lines.
}
\label{fig:2}
\end{figure}
In addition to scattering  charge impurities induce
strong carrier density inhomogeneities in graphene, especially close to the CNP,
that must be taken into account in the transport theory.
To characterize these inhomogeneities
we use the Thomas-Fermi-Dirac, TFD, theory \cite{rossi2008}
assuming that the impurities are
placed in a 2D plane at a distance $d=1$~nm from the graphene layer.
Fig.~\ref{fig:2}~(a), (b) show the carrier density profile
for a single disorder realization for the uncorrelated case and correlated case
($r_0=10\;a_0$) for $n_i =0.95\times10^{12}$~cm$^{-2}$. We can see
that in the correlated
case the amplitude of the density fluctuations is much smaller than in
the uncorrelated case.
The TFD approach is very efficient and allows the calculation of disorder averaged
quantities such as the density root mean square, $n_{\rm rms}$, and
the density probability
distribution $P(n)$. Figures \ref{fig:2}~(c), (d), (e) show $P(n)$ at
the CNP, and away
from the Dirac point ($n_i =0.95\times10^{12}$~cm$^{-2}$). In each
figure both the results for the uncorrelated case and the
one for correlated case are shown. $P(n)$ for the correlated case is in general
overall narrower than $P(n)$ for the correlated case resulting in smaller
values of $\nrms$ as shown in
Fig.~\ref{fig:2}~(f) in which $\nrms/n_i$ as a function of $r_0/r_i$ is plotted
for different values of the average density, $\nav$, and two different
values of the impurity density, $n_i =0.95\times10^{12}$~cm$^{-2}$
(``low impurity density'') for the
solid lines, and $n_i =4.8\times10^{12}$~cm$^{-2}$ (``high impurity
density'') for the dashed lines.

\begin{figure}
\includegraphics[width=0.99\columnwidth]{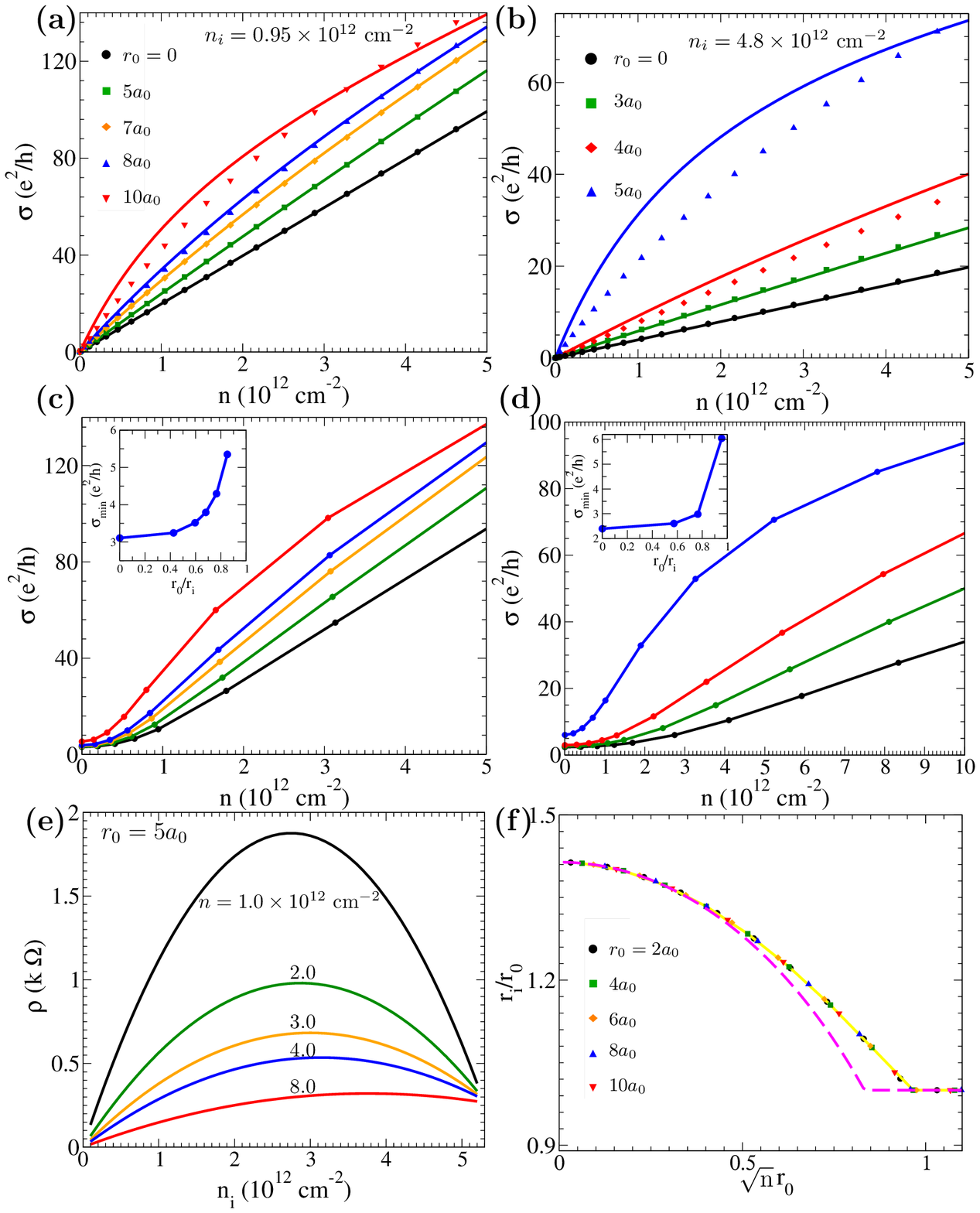}
\caption{
  Calculated $\sigma(n)$ with $S(\bq)$ obtained
  from the Monte Carlo
  simulations, symbols, and $S(\bq)$ given by Eq.~\ref{eq:strufac2}, solid lines for
  (a) $n_i=0.95\times 10^{12}$ cm$^{-2}$, and (b) $n_i=4.8\times 10^{12}$ cm$^{-2}$.
  (c) and (d) show the results for $\sigma(\nav)$ obtained from the EMT.
  The value used for $n_i$ in (c) and (d) is the same as in (a) and
  (b) respectively.
  The insets in (c) and (d) show the value of $\sigma_{min}$  as a
  function of $r_0/r_i$. In (e), the resistivity $\rho$ is shown as a
  function of impurity density $n_i$ for different carrier densities
  with $r_0 =5 a_0$. (f) The relationship between $r_i/r_0$
  and $\sqrt{n} r_0$ where the conductivity is minimum. The dashed
  line is obtained using Eq. \ref{eq:optimal}.
}
\label{fig:3}
\end{figure}

We now present our results for the conductivity.
The integral in Eq.~(\ref{eq:relatime}) can be calculated analytically for ``small"
$k_F$ by expanding $S(x)$ in the integrand giving:
\begin{equation}
\sigma(n)= A n \left[1 - a + B a^2 n/n_i\right]^{-1}
\label{eq:sigasym}
\end{equation}
where,
$ A  =   \dfrac{e^{2}}{h}\left[2n_{i}r_{s}^{2}G_{1}(r_{s})\right]^{-1} $,
$ a  =  \pi n_i r_0^2$, and
$ B  =  G_2 (r_s)/\left(2G_1(r_s)\right)$.
Note $a<1$ in our model.
The dimensionless functions $G_{1,2}(r_s)$ are given by,
$ G_{1}(x)  = \frac{\pi}{4}+6x-6\pi x^{2}+ 4x(6x^{2}-1)g(x)$, and
$G_{2}(x)  = \frac{\pi}{16}-\frac{4x}{3}+3\pi x^2
         +  40x^3  [1-\pi x+ \frac{4}{5} (5x^{2}-1)g(x) ]$,
where $g(x) = \text{sech}^{-1}(2x)/\sqrt{1-4x^2}$ for $ x<\frac{1}{2}$
and $\text{sec}^{-1}(2x)/{\sqrt{4x^{2}-1}}$ for $x>\frac{1}{2}$.
Eq.~(\ref{eq:sigasym}) indicates that for small $n$,
$\sigma(n)\sim An(1-a)^{-1}$, and for large $n$,
$\sigma(n)\sim\left(1-n_c/n\right)$ where $n_c=(1-a)n_i/(Ba^2)\sim
O(1/n_ir_0^4)$. The crossover density $n_c$, where the sublinearity
($n>n_c$) manifests itself, increases strongly with decreasing
$r_0$. This generally implies that the higher mobility annealed
samples should manifest stronger nonlinearity in $\sigma(n)$, since
annealing leads to stronger impurity correlations (and hence larger
$r_0$). This is exactly the experimental observation. While the
resistivity within the standard random model  increases linearly in $n_i$,
Eq.~(\ref{eq:sigasym}) indicates that the resistivity could decrease
with increasing impurity density if there is sufficient inter-impurity
correlations present in the system.
This is due to the fact that, for fixed $r_0$,
higher density of impurities are more correlated
causing $S({\bf q})$ to be more strongly suppressed at low $q$.
This is easy to see in the case in which
$r_0=a_0$ and $n_i$ so high that $r_i=r_0$. In this extreme case
the charge impurity distribution would be very correlated, indeed perfectly periodic, and
the resistance, neglecting other scattering sources, would be zero.
For each value of $r_0$ and
carrier density $n$, the maximum resistivity is found to be at
\begin{equation}
r_i/r_0 = \sqrt{2(1-\pi B n r_0^2)}.
\label{eq:optimal}
\end{equation}

In Figs.~\ref{fig:3}(a) and (b), we show  calculated $\sigma(n)$
using different values of the
impurity correlation parameters ($r_0$)  and $S(\bq)$ given by
Eq.~(\ref{eq:strufac2}) and Monte Carlo simulations.
The comparison between the
two results shows that the analytic continuum correlation model is
qualitatively and quantitatively reliable.
It is clear that, {\it for the same value of $r_0$}, the
dirtier (cleaner) system shows stronger nonlinearity (linearity) in a
fixed density range consistent with experimental observation since the larger impurity density $n_i$ of the dirtier system allows, in
principle, for stronger correlation effects to manifest itself due to
the fact that the crossover density $n_c$ is smaller for larger $n_i$.
To describe the transport properties close to the CNP and take into
account the strong disorder-induced carrier density inhomogeneities we
use the effective medium theory (EMT) \cite{Rossi}.
Fig.~\ref{fig:3}~(c) and (d) show the EMT results for $\sigma(n)$.
The insets in Fig.~\ref{fig:3}~(c) and (d) show the dependence
of $\sigma_{mim}$ on the size of the correlation length $r_0$.
$\sigma_{mim}$
increases slowly with $r_0$ for $r_0/r_i<0.5$, but quite rapidly
for $r_0/r_i>0.5$.
Finally, Fig. \ref{fig:3}(e) shows that the resistivity ($1/\sigma$) is highly
nonlinear as a function of impurity density and the optimal $r_i/r_0$ at
which the conductivity is minimum [Fig. \ref{fig:3} (f))].
%
%
%

%

%
The results shown in Fig.~\ref{fig:3} strikingly demonstrate the full
power of the impurity
correlation model as it clearly produces the observed experimental
behavior with strong sublinear behavior for stronger impurity
correlations (i.e. larger $r_0$). Annealing leads
to stronger correlations among the impurities since the impurities can
move around to locate to equilibrium sites, thus enhancing $r_0$,
which strongly suppress the crossover carrier density $n_c(\sim
r_0^{-4})$, thus increasing the overall nonlinearity of
$\sigma(n)$.
In addition the theory
explains the observed strong nonlinear $\sigma(n)$ in suspended
graphene \cite{BolotinYacoby} where the thermal/current annealing is
used routinely.
Finally, graphene on hexagonal BN is likely to have significant
correlations in the impurity locations imposed by the similarity
between graphene and BN lattice structure. This implies stronger
nonlinearity in the $\sigma(n)$ dependence for graphene/BN system as
has recently been observed experimentally \cite{BNDeanDas}. Although we have used a minimal model for impurity correlations, using a single correlation length parameter $r_0$, which captures the essential physics of correlated impurity scattering, it should be straightforward to improve the model with more sophisticated correlation models if experimental information on impurity correlations becomes available \cite{fuhrer2010}.

In summary, we provide a novel physically motivated explanation for
the observed nonlinear behavior of graphene conductivity
by showing that the inclusion of spatial
correlations among the charged impurity locations leads to a
significant sublinear density dependence in the conductivity in
contrast to the strictly linear-in-density graphene conductivity for
uncorrelated random charged impurity scattering. The great merit of
our theory is that it  eliminates the need for an {\it ad hoc}
zero-range defect scattering mechanism which has always been used in
the standard model of graphene transport in order to
phenomenologically explain the high-density sublinear behavior. Even though the
short range disorder is not needed to explain the sublinear behavior
in our model we do not exclude the possibility of short range
disorder scattering in real graphene samples, which would just add as
another resistive channel with constant conductivity. We mention that a recent experimental work\cite{fuhrer2010} reports graphene transport data in remarkable agreement with the theory developed herein.

{\it Acknowledgements} This work is supported by ONR-MURI and NRI-SWAN.
ER acknowledges support from the Jeffress Memorial Trust, Grant No. J-1033. Computations were carried out in part on the SciClone Cluster at the College of William and Mary. We thank Michael S. Fuhrer and Jun Yan for discussions and for sharing with us
their unpublished experimental data.

\end{document}